\documentstyle[11pt]{article}
\newcommand{\epspr}{\epsilon^\prime}
\begin{document}
\begin{Large}
\begin{center}
\noindent In Search of a Source for the 320 EeV Fly's Eye Cosmic Ray
\end{center}
\end{Large}
\vspace{1cm}

\begin{center}
\Large{Jerome W. Elbert and Paul Sommers}\\
\vspace{1cm}
\begin{large}
High Energy Astrophysics Institute\\
Department of Physics, University of Utah\\
Salt Lake City, UT 84112, USA\\
\vspace{10mm}
\end{large}

\vspace{1cm}

\noindent \begin{em} A postscript file of figures is available via
anonymous ftp to

\noindent einstein.physics.utah.edu (get pub/search-figs.ps.Z).
\end{em}

\vspace{1cm}
\Large{ABSTRACT}
\end{center}

{\large The 320 EeV air shower detected by the Fly's Eye poses an
important problem.  Careful analysis of pathlength limitations
for the possible particle types due to cosmic background
radiation verifies that the particle very likely traveled less
than 50 Mpc from its source. The best candidates for accelerating
particles to such high energies are the very powerful
radiogalaxies, however they are all more than 100 Mpc distant.
Our search finds no likely source within 50 Mpc in the direction
from which the particle arrived.  This prompts consideration of
less likely astrophysical sources, like M82, as well as
non-standard mechanisms like cosmic string annihilation.  It is
also conceivable that the air shower was produced by some
non-standard particle whose pathlength is unlimited because it
does not interact with the cosmic background radiation.  A less
radical alternative is that relatively strong magnetic fields
deflected the particle's path through a large angle, so it could
have originated at a nearby radiogalaxy at an earlier time of
strong activity.} ~~\\ ~~\\
\begin{large} \begin{em} Subject headings: \end{em} acceleration of
particles -- cosmic rays -- cosmic strings -- galaxies: magnetic fields --
elementary particles -- galaxies: intergalactic medium

\end{large}

\newpage
\vspace{10mm}

\begin{large}
\section{Introduction}

	Discovering the origins of the highest energy cosmic rays
has been a goal in astroparticle physics for three decades since
the first reported detection of a particle with energy greater
than 100 EeV ($10^{20}$ eV) (Linsley 1963).  The Fly's Eye air
shower with energy $320\pm 93$ EeV should be a useful clue (Bird
et al. 1994).  This energy is substantially higher than that of
any previously reported cosmic ray, and the issues concerning its
origin are much more sharply focused.  At such high energy,
particles cannot propagate long distances through the cosmic
background radiation.  As will be discussed in detail in section
2, the pathlength is almost surely limited to less than 50 Mpc.
The magnetic rigidity of such a particle, moreover, is large
enough that it should be deflected little by the Galaxy's
magnetic disk or by the smaller extragalactic fields which may
exist between us and the source.  In that case, the arrival
direction of the air shower should point approximately toward its
source, and the source should be a prominent astrophysical object
within a distance of 50 Mpc.

	There is, unfortuneately, no likely astrophysical source
for such a high energy particle near its arrival direction and
within 50 Mpc.  Details of our search for a candidate source will
be presented in section 3.  Ordinary galaxies are not expected to
accelerate particles to such high energy.  Hillas (1984) analyzed
the source requirements.  Acceleration models generally require
magnetic media moving with high relative velocity, and the models
need a large value for the product of magnetic field strength
times the characteristic size of the astrophysical system.  Giant
radiogalaxy hot spots most nearly meet the requirements.  Rachen
and Biermann (1993) have argued that some hot spots may indeed
possess conditions which would produce such energetic particles
by first-order Fermi acceleration.  Acceleration at an accretion
shock of a massive black hole in an active galactic nucleus is
{\em not} a viable model because of the high photon densities
there and the consequent rapid energy losses due to collisions
with them (Stecker et al. 1991).  It is conceivable, however,
that an explosion in a galactic nucleus could cause a strong
magnetic shock at a larger radius where photon densities are low
enough to permit the acceleration.  In shock acceleration
scenarios, one might expect that there should also be accelerated
electrons which would produce strong radio emissions.  Our search
in section 3 therefore emphasizes extragalactic radio sources
near the arrival direction of the Fly's Eye air shower.

	The arrival direction of the Fly's Eye air shower in 1950
celestial coordinates is ($\alpha = 85.2^{\circ} \pm
0.5^{\circ}$, $\delta = 48.0^{\circ} \pm 6^{\circ}$).  The
uncertainty in right ascension is small because the
track-detector plane nearly contains the celestial poles (Bird et
al. 1994).  The directional uncertainty within that plane
produces the larger uncertainty in declination.  In galactic
coordinates, the error box is centered on ($b=9.6^{\circ},\
l=163.4^{\circ}$).

	Since the arrival direction of the 320 EeV particle does
not point to any obvious source within the distance limit, there
may be some new physics or astrophysics to be learned from
it.  Some hypotheses to be considered in this paper are the
following:

	(1) Perhaps there are larger-than-expected magnetic
fields outside the Galaxy.  It is then possible that the
particle's trajectory was bent through a large angle.  It is
important to note that its pathlength would then be much longer
than that of photons with which we observe astrophysical sources.
For example, the radiogalaxy Centaurus A is at a distance of
about 3 Mpc (Hesser et al. 1984), but makes an angle of
136$^{\circ}$ with the arrival direction of the 320 EeV particle.
For a trajectory of constant curvature, the particle would have a
pathlength of 10 Mpc, which means it was produced more than 20
million years prior to the emission of the photons which
constitute our current picture of Cen A.  That time is longer
than the decay time for strong radio galaxy activity (Schmidt
1966; van der Laan \& Perola 1969).  It is likely that Cen A was
indeed extremely active in the past (Hey 1983).  It has two sets of
radio lobes (Clarke, Burns \& Norman 1992), and the separation of
the outer set is as great as in the strongest radiogalaxies like
Cygnus A (Dreher, Carilli \& Perley 1987).  Cen A is not the only
candidate source if such extragalactic magnetic
trajectory bending is possible.  Virgo A, whose direction
makes an angle of $87^{\circ}$ with the arrival direction of this
particle and whose distance is 13-26 Mpc, also has a double-lobe
structure on a much larger scale than that of the jet which is
currently active (Kotanyi 1980).

	(2) Less magnetic bending is needed if the particle
originated at the starburst galaxy M82, whose direction differs
only by $37^{\circ}$ from the particle's arrival direction and
which is only about 3.5 Mpc away.  (The distance to M81, a member
of a group of galaxies which includes M82, has been discussed
recently by van den Bergh (1992), de Vaucouleurs (1993), and
Madore, Freedman, and Lee (1993).)  If the energy of this shower
was actually near 230 EeV (near the experimental $1\sigma$ lower
limit), then the shower's primary particle could have been a
nucleus of $Z\sim 18$ surviving from an iron nucleus accelerated
to about 340 EeV at M82.  The magnetic rigidity of a highly
charged nucleus would be low enough that the deflection through
37$^{\circ}$ could be achieved by our Galaxy's magnetic disk.
Also, the acceleration to superhigh energy is easier for a highly
charged nucleus (like iron) than for a proton. Strong magnetic
shocks may be associated with the intense starburst activity and
superwind, and such shocks might accelerate nuclei to superhigh
energies.  The high density of expanding supernova shells in the
starburst regions might also produce superhigh energy nuclei.

	(3) Although there are no nearby likely sources near the
direction of the detected air shower, there are some interesting
objects near that direction at greater distances.  One of these
is the Seyfert galaxy MCG 8-11-11, also known as UGC 03374,
which is at a distance of
62-124 Mpc (for Hubble constant in the range 50-100 km s$^{-1}$ Mpc$^{-1}$).
It has a strong luminosity of $7\times 10^{46}$ erg/s in low
energy gamma rays.  A comparable luminosity above 320 EeV would
be more than enough to account for the Fly's Eye detection,
despite the attenuation factor of 0.0028 associated with
propagating 62 Mpc.  This picture raises a question, however, as
to why there should not be a large flux of particles from roughly
the same direction at lower energy (10-50 EeV), where attenuation
would be insignificant.

	(4) One way to avoid that question is to conjecture that
there is some type of particle which is immune to interactions
with the cosmic background radiation, but which produces a normal
air shower in the atmosphere.  For example, perhaps neutrinos
have a much higher interaction cross section than expected at
such high energies.  Or perhaps the shower was produced by some
other non-standard neutral particle.  Such conjectures can avoid
the distance limits.  It is then interesting to note that there
is a remarkable quasar (3C147) within the nominal error box for
the arrival direction of this shower.  That source has a radio
luminosity more than 2000 times greater than Cen A or Virgo A,
and there is observational evidence for strong magnetic fields
there.  Even with no attenuation, however, its distance is so
great that the luminosity required above 320 EeV probably exceeds
the total luminosity of the quasar.

	(5) The puzzle of the missing source can be resolved
alternatively by supposing that the particle was not accelerated
at any persistent astrophysical object, but originated instead
from the annihilation of some topological defect like a cosmic
string (Aharonian, Bhattacharjee, \& Schramm 1992; MacGibbon \&
Brandenberger 1993).  Such annihilations are expected to result
in the production of GUT-mass particles ($mc^2 \approx 10^{24}\
eV$) which decay to superhigh energy hadrons.  The superhigh
energy flux reaching Earth might be dominated by neutrinos and
gamma rays resulting from pion decays.  The Fly's Eye measurement
of the shower's longitudinal profile does not exclude a gamma ray
as the primary particle.  A cosmic string annihilation is a
viable scenario.

	We assume throughout this paper that the 320-EeV particle
did not originate in the Galaxy.  This is partly because it is
not expected that conditions exist anywhere in the Galaxy for
accelerating particles to such high energy, based on the
arguments of Hillas (1984).  In addition, the Fly's Eye
composition and anisotropy studies have suggested that the
highest energy cosmic rays do not originate in the Galaxy's disk
(Bird et al. 1993a).  Furthermore, the arrival direction of this
shower does not point toward the Galactic center or any prominent
galactic source of high energy radiation.  It is more than
20$^{\circ}$ from either the Crab nebula or Geminga.  A galactic
origin for this particle cannot be rigorously excluded, however.
Its arrival direction is less than 10$^{\circ}$ from the Galactic
plane and is near to one of the clusters of showers above 10 EeV
which Chi et al. (1992) suggest are due to galactic sources.  It
should be noted, however, that 18\% of the Fly's Eye's exposure
is within 10$^{\circ}$ of the galactic plane, so the equatorial
latitude of this shower could be an accident.  Also, tracing
backward the trajectory of a positively charged particle with
this arrival direction through the Galaxy's regular magnetic
field, it is observed to bend northward away from the Galactic
plane rather than curving back toward any source at the plane.
Acceleration in an extended halo of the Galaxy cannot be ruled
out either.  In the Jokipii and Morfill (1985) model, however,
particles do not reach 320 EeV per nucleon, and heavy nuclei
photodisintegrate before reaching 320 EeV total energy.  The
possibility of a galactic origin for this particle is not pursued
in this paper.  The source of the extraordinary Fly's Eye shower
is assumed to be extragalactic.

\section{Distance limits}

\subsection{Nucleons}

	At 320 EeV, a proton or neutron loses energy primarily
through pion photoproduction:
\begin{equation} p\gamma \Rightarrow p\pi^0
\hspace{1.5cm}or\hspace{1.5cm} p\gamma \Rightarrow n\pi^+
\end{equation}
\begin{equation} n\gamma \Rightarrow n\pi^0 \hspace{1.5cm}or\hspace{1.5cm}
n\gamma \Rightarrow p\pi^-. \end{equation}  The charge exchange reactions are
approximately as common as those in which the nucleon isospin
does not change.  The mean free path is approximately 5 Mpc, and
the emerging nucleon typically gets 80\% of the incoming
nucleon's energy.  The 2.73$^{\circ}$ microwave photons constitute the
main targets.  The neutron decay mean free path (3 Mpc) is less
than the collision mean free path, so a nucleon
spends the majority of its time as a proton.  At this energy, the
energy loss rate due to $e^{\pm}$ pair production (Blumenthal 1970) is
negligible compared to the pion photoproduction losses.

	We have analyzed the effects of pion photoproduction on
nucleon propagation using Monte Carlo programs which follow the
history of a nucleon through successive interactions until its energy
has been reduced below a threshold value.  A brief account of the
Monte Carlo method is given here.  We begin by considering a nucleon
of energy $E=\gamma mc^2$ moving through an isotropic intensity of
photons of energy $\varepsilon$.  Denoting by $\theta$ the angle a
photon's direction makes with the nucleon's direction, the effective
interaction cross section is \begin{equation} \sigma
_{eff}(\varepsilon)=\frac{1}{4\pi}\int_{0}^{\pi}2\pi sin \theta (1-cos
\theta)\sigma (\varepsilon')d\theta \equiv
\frac{1}{2\gamma^2\varepsilon^2} \int_0^{2\gamma \varepsilon}
\varepsilon' \sigma (\varepsilon')d\varepsilon', \end{equation} where
$\varepsilon'
\equiv \gamma \varepsilon (1-cos \theta)$ is the photon's energy in
the nucleon's restframe, and $\sigma (\varepsilon')$ is the total
interaction cross section for a $\gamma N$ interaction as a function
of $\gamma$-ray energy $\varepsilon'$.  In the $2.73^{\circ}$K
blackbody radiation, the (inverse) mean free path is gotten by
integrating over the photon energies:
\begin{equation}
\lambda^{-1}=\int_0^{\infty}\frac{dn}{d\varepsilon}\sigma_{eff}
(\varepsilon)d\varepsilon = \frac{1}{2\pi^2\hbar^3c^3\gamma^2}
\int_0^{\infty}\frac{1}{e^{\varepsilon/kT}-1}\{\int_0^{2\gamma\varepsilon}\varepsilon'
\sigma(\varepsilon') d\varepsilon' \}d\varepsilon. \end{equation} For
$\sigma(\varepsilon')$, we use the total cross section shown in Figure
2 of Hill \& Schramm (1985).  The mean free path $\lambda$ defines an
exponential pathlength distribution from which the Monte Carlo program
samples an interaction step.  The $\gamma$-ray energy $\varepsilon'$
is sampled from the distribution implicit in the above integration.
The center-of-momentum energy is thereby determined, and the nucleon's
outgoing energy is determined once its direction in the
center-of-momentum frame is chosen from an isotropic distribution
(Hill \& Schramm 1985).

	A Monte Carlo simulation was used to evaluate the changes in a
proton spectrum after propagation through various fixed distances
ranging from 3 Mpc to 100 Mpc. Protons were sampled from a power law
spectrum with a differential spectral index $\gamma =$2.5.  The
sampled energies ranged from 10 EeV to 10$^4$ EeV.  Results are given
in Figure 1a.  The vertical axis gives the resulting {\em integral}
spectrum multiplied by $E^{\gamma-1}$ and the vertical scale is
arbitrarily normalized to unity for an unmodified spectrum.  Separate
curves are plotted for the different propagation distances. The
effects on the spectrum are minimal below about 30 EeV, but become
significant near 100 EeV.  The attenuation is more severe for greater
path lengths and the spectrum is reduced by a factor of about 0.005 for
the integral flux above 320 EeV after a pathlength of 50 Mpc.

	Of special interest here is the fraction of particles
produced above 320 EeV which remain above that energy after
various propagation distances.  In Figure 2a the surviving
fraction of the integral flux above 320 EeV is plotted as a
function of propagation distance.  The solid, dashed, and dotted
lines represent results for $\gamma=$ 2.0, 2.5, and 3.0,
respectively.  All curves show attenuation by more than or
approximately 2 orders of magnitude for a propagation distance of
50 Mpc. The most penetrating spectrum is that with $\gamma =
2.0$.  Compared with the other spectral indices, the mean proton
energy at production is highest for $\gamma=2.0$ and the mean
free path is {\em shortest}, but the higher energy allows a
larger number of interactions to occur before the proton energy
falls below 320 Eev.  The second factor (more interactions
possible before dropping below 320 EeV) dominates, so that the
``hardest'' spectrum is the most penetrating. Roughly speaking,
however, all three spectra are attenuated approximately
exponentially, with attenuation lengths near 10 Mpc.

    We can use figure 2a to derive a ``$2\sigma$ distance
limit.''  In order to do this, we assume that the source of the
320 EeV shower has a production spectrum no flatter than a power
law with differential spectral index 2.  We also assume that
magnetic deflection from the source direction is not greater than
about $30^{\circ}$ for protons of 10 EeV and above.  All nucleons
above 10 EeV from the source should then arrive within a $2\pi/3\
sr$ sky region centered on the galactic anticenter direction.  It
has been reported (Bird et al. 1993b) that the Fly's Eye detected
71 showers above 10 EeV from that region of the sky.  Allowing
for a $2\sigma$ downward fluctuation in the detected flux, the
expected number above 10 EeV from the source cannot be
greater than 90, and pion photoproduction attenuation is
negligible at 10 EeV.  Using an E$^{-1}$ dependence for the
integral flux gives 2.81 as the upper limit for the expected
number of showers above 320 EeV from the source if there were no
attenuation.  We assume that the probability of detecting at
least one shower above 320 EeV from the source was not less than
0.0454 (the $2\sigma$ probability level).  Let F(D) be the
integral flux reduction factor plotted in Figure 2a.  Then $
2.81\times F(D) \geq 0.0454\ \Rightarrow \ F(D)\geq 0.0162. $
{}From Figure 2a, it is seen that this relation is satisfied only
if D$\leq$47 Mpc.  Therefore, 47 Mpc is a $2\sigma$ distance
limit for nucleons.  (This argument implicitly uses the fact that
the Fly's Eye acceptance is almost independent of energy above 10
EeV.)  For the source search of section 3, we round off this
distance limit to 50 Mpc.

	An example may clarify the reasoning which leads to this
$2\sigma$ distance limit.  Suppose the 320 EeV particle came from
the Seyfert galaxy MCG 8-11-11.  For a Hubble constant of 100
km s$^{-1}$ Mpc$^{-1}$, its distance is 62 Mpc.  The integral flux reduction
factor at 62 Mpc (from Figure 2a) is $F(62)=2.8\times 10^{-3}$.
We assume that the Fly's Eye was not extraordinarily lucky to
detect that shower, i.e. the probability was at least as great as
the $2\sigma$ probability 0.0454.  In order to have a probability
of at least 0.0454 for detecting a shower above 320 EeV after an
integral flux reduction by $2.8\times 10^{-3}$, the expected
number of detected particles without attenuation would have to be
at least 15.1 (i.e., 0.0454/0.0028).  Then, under
the assumption that its integral production spectrum is no
flatter than E$^{-1}$, we find that the expected number is at
least 483 above 10 EeV (where attenuation is negligible).  This
exceeds the experimental $2\sigma$ upper limit of 90 events from
all sources in the $2\pi/3\ sr$ sky region which includes MCG
8-11-11.  (In fact, it exceeds the number observed from {\em all}
parts of the sky above 10 EeV.)  In brief, MCG 8-11-11 is far
enough away that detecting even a single shower from its
attenuated flux above 320 EeV would imply (assuming an E$^{-1}$
or softer integral source spectrum) a detectable flux above 10
EeV in excess of the observational upper limit.  MCG 8-11-11 is
therefore beyond the $2\sigma$ distance limit.

	 There is an alternative method for evaluating a distance
limit which does not require any assumption about the source
spectrum or any evaluation of what the observed flux would be in
the absence of attenuation by pion photoproduction.  Instead, it
relies on an assumption that sources may occur anywhere and are
located uniformly in space.  Then we can ask for the probability
that a particle with E$\ge 320$ EeV traveled a path length D or
greater before its detection.  This amounts to asking, for the
totality of particles above 320 EeV, ``What is the fraction of
the time spent at path lengths greater than D from the production
sites?''  The dependence on D of this probability is plotted in
Figure 3a, where the different curves represent different
spectral indices as in Fig. 2a.  The probability that the source
is beyond 50 Mpc is less than 0.007 in all three cases.  The
assumption of a spatially uniform distribution of possible
sources may be appropriate for the model of topological defect
annihilations.  It is also an interesting approximation if
superhigh energy particles can originate in normal galaxies or
other common astrophysical entities.  Like Figure 2a, Figure 3a
shows an approximately exponential dependence on distance with
attenuation length near 10 Mpc.  The interpretation that we give
to this result is that, without specific knowledge about the
sites of superhigh energy particle production, the best guess is
that the particle traversed a pathlength on the order of 10 Mpc.

\subsection{Nuclei}

\par Collisions of Universal Microwave Radiation (UMR) photons with
 high energy nuclei are important in the energy range near 320 EeV.
 The importance of these effects for the propagation of energetic
 nuclei was noted by Greisen (1966) and by Zatsepin and Kuz'min
 (1966). Stecker (1969) calculated characteristic lifetimes for
 $^4$He and Fe nuclei in the UMR and found that the
 photodisintegration effects begin to produce significant limitations
 (mean lifetimes $\tau \ll 10^{10} $ yr) in the energy range 10 EeV
 (for He) to 100 EeV (for Fe).

\par  Based on equation 4, we have performed simulations of the
 propagation of various energetic nuclei. In developing the Monte
 Carlo programs it was necessary, for incident nuclei of nuclear mass
 i and fragments of nuclear mass j, to find appropriate
 photo-spallation cross sections $\sigma_{ij} (\epspr )$ for all
 $\epspr$ (nuclear rest frame gamma ray energies) of interest and for
 all $i \leq 56$ and all $j < i$.  The simulations were done for
 nuclei with initial energies ranging from 10 to 10$^4$ EeV. For
 $\epspr \leq$150 MeV, these cross sections
 are available from Puget, Stecker, and Bredekamp (1976).

 \par However, $\epspr > $150 MeV sometimes occurs in the present
simulations. For example, a head-on collision of a UMR photon
with a typical energy $\epsilon = 6 \times 10^{-4}$ eV gives
$\epspr =$ 3.2 GeV in the center of mass system of a $10^4$ EeV
$^4$He nucleus. Consequently it was necessary to estimate higher
energy cross sections than are available in Puget et al. (1976).

 \par Several fitted approximations to photonuclear spallation cross
 sections have been done which extend into the GeV energy range.  A
 number of approximations done for  incident protons and gamma rays
 are modifications of the form given by Rudstam (1966). Although
 Rudstam fitted data for incident protons, Jonsson and Lindgren (1973,
 1977) used a related 5 parameter formula for photo-spallation data.
 They pointed out similarities between three of their parameters and
 those of Rudstam, but they noted differences between the cases of
 incident gamma rays and incident protons. The most important
 difference was in the normalization of the cross sections.
 Jonsson and Lindgren (1973) approximated the total cross section
 factor by $\hat{\sigma}=0.3 A_t$ (mb) for photo-spallation, where
 $A_t$ is the mass number of the target nucleus, while Rudstam gave
 $\hat{\sigma}=50 A_t^{2/3}$ (mb) for proton-induced spallation.

\par Silberberg and Tsao (1973a,b) give extensive and detailed fits to
 nuclear spallation data for incident protons. We have used this
 model with a renormalization factor given by the ratio of the above
$\hat{\sigma}$ approximations for incident photons and protons. The
 factor is $0.006 A_t^{1/3}$.  The model parameters were taken from
 Tables 1-A to 1-D of Silberberg and Tsao (1973a), except that the
 formulas for peripheral reactions (Silberberg and Tsao 1973b) were
 used for photo-production of a single proton, or one to three
 neutrons, or a single proton and a number of neutrons (fewer than
 some specified number which depends on the target nucleus mass.)
 The resulting cross sections are in reasonable agreement with
 photospallation data for the production of $^{24}$Na on nuclei with
 10 A$_t$ values ranging from 27 to 64  shown in Fig. 7 of Jonsson
 and Lindgren (1973). A comparison of the discontinuities at 150 MeV
 between the cross sections obtained by this method and those of
 Puget et al. (1976) show rms deviations of about a factor of 3.
 The cross sections described here were used only for $\epspr > $150
 MeV and, considering the discrepancies noted above, are regarded
 only as estimates in the GeV energy range.
\par Equation 4 can be used to give the mean free path for nuclear
 spallation in a single radiation field.  We have followed
 Puget et al. (1976) in including contributions from microwave,
 optical, and infrared radiation.  The microwave radiation was
 evaluated for a 2.73 K black body. Following Puget et al. (1976), the
 intergalactic optical spectrum was estimated assuming a 5000 K
 black body, whose photon density is reduced by a factor of
 $1.2 \times 10^{-15}$.  The infrared spectrum was taken from
 De Jager, Stecker, and Salamon (1994)
 who found $n(\epsilon )=0.6 \pm 0.3 \times 10^{-3}
 \epsilon ^{-2.6}(H_0 / 75$ km s$^{-1}$ Mpc$^{-1}$) $cm^{-3}eV^{-1}$.
 The infrared spectrum is used for photon energies from 0.002 eV to
 0.8 eV. Besides spallation, the nuclei lose energy by pair
 production.  We have included the pair production loss rates given
 by Puget et al. (1976) in a continuous energy loss approximation
 in the Monte Carlo simulations.

 For nuclei starting out with masses A=12 (C) and A=56 (Fe),
 results were obtained which are displayed in Figures 1-3 along
 with results for protons which were discussed in the previous
 section.  For all the plots involving nuclei, the energies are the
 total energies per remaining nucleus, and fragments with A $<$ 3 are
 ignored. Figures 1b (for C) and 1c (for Fe) show the fraction of the
 integral spectrum remaining after nuclei have traveled various
 distances.  There are some similarities in shape between these
 curves and those from protons, but the differences are important.
 The flux reduction is more severe for nuclei than for protons.
 For energies above 200 EeV, both C and Fe are severely attenuated
 for distances above 10 Mpc. Below 100 EeV, the attenuation is more
 severe for C than for Fe, although for 3 and 10 Mpc C is more
 penetrating than Fe at 320 EeV.  As energy increases for both the
 C and Fe curves at 10 Mpc there is a sharp fall followed by a rapid
 transition to a much flatter region.  In these all-nuclei spectra,
 these flatter regions correspond to the emergence of the spectra of
 lower A, higher $\gamma$ fragments which have weaker energy
 dependence of the nucleon loss rates than the initial nuclei.

 \par In Fig. 2b (for C) and Fig. 2c (for Fe) the surviving fractions
 of the integral spectra above 320 EeV are shown for 3 spectral
 indices.  If we approximate these curves by exponentials (not a very
 good approximation for Fe at D $<$ 3 Mpc), the characteristic
 lengths are near 1.5 Mpc for C and on this order of magnitude for
 Fe. The distributions of path length probabilities (defined in the
 previous section) are shown in Fig. 3b (for C) and Fig. 3c (for Fe).
 The results are numerically similar to those of Figures 2b and 2c.
 Although only results for C and Fe are displayed here, all A values
 from 3 to 56 have been simulated in the Monte Carlo program and none
 have propagation characteristics which are much more favorable that
 the examples used here. The main conclusion of this section is that
 for path lengths greater than 10 Mpc heavy nuclei are not attractive
 candidates for the 320 EeV primary particle except in models in
 which it is acceptable to have severe attenuation between the point
 of production and Earth.

\subsection{Gamma rays}

	The propagation of gamma rays through cosmic radiation
fields has been studied in detail by others (Gould \& Schreder
1967; Wdowczyk, Tkaczyk, \& Wolfendale 1972; Halzen et al. 1990).
The mean free path between collisions with microwave photons at
this energy is 58 Mpc.  More important is the radio photon
spectrum.  There is some uncertainty about the cosmic radio
spectrum, but using the spectrum of Clark, Brown, and Alexander
(1970), the mean free path for 320 EeV gamma rays is 7.4 Mpc.  For
radio and microwave photons together, the mean free path is 6.6
Mpc.  It has been suggested that the leading electron produced
in a $\gamma
\gamma$ collision could deliver most of the original gamma ray's
energy to a new gamma ray via inverse Compton scattering.  If
intergalactic magnetic field strengths exceed $10^{-11}$ G,
however, then synchrotron losses will prevent this (Wdowczyk et
al. 1972).  We here assume that intergalactic fields are at least
that strong.  The survival probability for 320 EeV gamma rays to
distance D in Mpc is then given by \begin{equation}
P(>D)=exp(-D/6.6).\end{equation} This survival probability
becomes less than 1/1000 before reaching a distance of 50 Mpc.

\subsection{Neutrinos}

	The dominant attenuation of 320 EeV neutrinos is due to
interactions with the 1.9$^{\circ}$K blackbody neutrinos (Yoshida
1994), but the mean time between interactions far exceeds the
Hubble time.  So neutrinos can come from arbitrarily distant
sources.  It is implausible that the Fly's Eye air shower was
initiated by a neutrino, however, since the interaction
probability for such a neutrino should be less than $10^{-5}$
while traversing the entire atmosphere.  Moreover, the point of
first interaction should be equally likely at all atmospheric
depths, whereas this shower certainly started high in the
atmosphere.

	On the other hand, the interaction cross section for
superhigh energy neutrinos is not experimentally confirmed.
Expected cross sections are calculated with the Standard Model,
and it is conceivable that new physics at high energies might
enhance neutrino cross sections (e.g. Domokos \& Nussinov 1987;
Domokos \& Kovesi-Domokos 1988).  If the cross section were as
large as hadronic cross sections, then such neutrinos would
interact high in the atmosphere, producing air shower
developments which could not be distinguished from other cosmic
ray air showers on the basis of Fly's Eye data.  Searches have
been made for deeply penetrating particles in the Fly's Eye data
without success (Baltrusaitis et al. 1985; Emerson 1992).  Based
on the standard cross section calculations, this has been
interpreted as a weak limit on the intensity of neutrinos above
0.1 EeV.  If the neutrino interaction cross section is large at
all energies above 0.1 EeV, however, the Fly's Eye would not be
able to distinguish the neutrino intensity from other cosmic
rays.  This non-standard hypothesis for a large neutrino
interaction cross section at extremely high energies offers a
speculative way to evade the distance limits which pertain to the
other particle types.

\section{The source search}

\par An attempt has been made to identify plausible astronomical
 objects which could have produced the 320 EeV primary particle.
As mentioned previously, we believe that the highest energy Fly's
Eye showers originate outside the Galaxy. If we exclude the
Galaxy, it is reasonable to assume that the source is a galaxy
which differs from our Galaxy and similar nearby galaxies in
having unusual luminosity in at least some part of the
electromagnetic spectrum.  The galaxies of interest should be
strong radio sources if magnetic fields and large fluxes of
energetic particles are present. In the previous section we
showed that the ranges of nucleons, nuclei, and gamma rays are
less than or on the order of 50 Mpc. Because the distances of
interest are small relative to typical distances in radio galaxy
catalogs, the number of potential sources is not large and a
powerful galaxy in the volume of interest should not have been
missed.

\par If we assume that extragalactic magnetic fields are small (less
 than about $ 10^{-9}$ G), then the primary particle would be
deflected by less than about $10^\circ$ from its original source
direction. In this case the source direction is probably within
$10^\circ$ of the boundaries of a 2$\sigma$ error box drawn
around the shower's reconstructed arrival direction. This region
is represented by the outer boundary in Fig. 4. We have searched
for potential sources using a catalog of 459 powerful radio
galaxies with measured redshifts (Burbidge and Crowne 1979). The
galaxies included in this catalog were selected with a cut on
radio luminosities ($L > 10^{41}$ ergs $s^{-1}$.) This cut is
appropriate because, as stated by Burbidge and Crowne (1979),
``Weak or normal radio galaxies have luminosities in the range
$10^{37}-10^{40}$ ergs s$^{-1}$. By using the luminosity
criterion given above, we have deliberately omitted from the
catalog the large number of comparatively bright nearby galaxies
which are weak sources similar to our own galaxy.''  In this
paper, we also make a cut on the redshift parameter, $z$, which
requires $z<0.0125$ which corresponds, for a Hubble constant of
75 km s$^{-1}$ Mpc$^{-1}$, to distances less than 50 Mpc. With
the redshift cut, but without restricting the directions, 24
candidate galaxies remain, including such well-known galaxies as
For A, M82, M84, M87 (Vir A), and Cen A. However, none of the
candidate galaxies are within the outer boundary of Fig. 4.

\par A second search was done using the NASA/IPAC Extragalactic
 Database (NED) (G. Helou et al. 1991).  The goal of this search was
 similar to that described in the previous paragraph, except that the
 search was limited to the region inside the outer boundary in
 Fig. 4. In order to make a cut on the radio luminosity using data
 taken at various frequencies, a criterion was defined which required
 the radio luminosity per decade of frequency to exceed
 $10^{40}$ ergs s$^{-1}$.  This requires that the measured flux
 density, $S$, in Jansky, exceeds $S_{min}$:
 \begin {equation} S_{min} = \frac{2.3 \times 10^5} {\nu z^2}.
 \end {equation} The frequency, $\nu$, is given in Hz and $z$ is the
 redshift parameter of the source. This cut separates sources of
 exceptional radio luminosity from ordinary sources. We can define a
 dimensionless luminosity parameter $R_L \equiv S / S_{min}$.
 (Where multiple radio measurements have been made on a source, the
 maximum value of $R_L$ is used.) Some examples illustrate the range
 of the luminosity parameter. For galaxies which are not outstanding
 in radio emission, such as M31 and M81, the parameter is equal to
 0.00002 and 0.002, respectively.  The nearby starburst galaxy M82
 has $R_L=0.017$. The prominent radio sources Cen A and M87 give
 $R_L$ values of 11. and 17., respectively, while 3C147 (discussed
 further below) gives  $4 \times 10^4$. With a cut on z ($z<0.0125$),
 but no $R_L$ cut, the NED search yielded one galaxy within the outer
 boundary of Fig. 4. The galaxy is NGC 1569, or Arp 210.
 However, $R_L$ is 0.0015 for this galaxy. Since the radio luminosity
 is so unremarkable, this galaxy is not regarded as a likely source.

\par A second NED search was done which was like the first except
 that the $z$ limits were enlarged to $z < 0.0148$, which corresponds
 to a distance of 47 Mpc for the case in which
 $H_0 = 100$ km s$^{-1}$ Mpc$^{-1}$.  This search yielded an additional
 candidate, UGC 03351 (or MCG 10-09-02).  Its direction is nearly on the
 shower-detector plane for the 320 EeV event, but it is nearly $2 \sigma$
 north of the best fit direction.  The direction is plotted in
 Fig. 4.  It is a spiral galaxy, type Sab.  The z value is near the
 limit of this search ($z = 0.0148$), and the radio luminosities at
 1.4 and 4.85 GHz both give $R_L$ near 0.2. The luminosity of this
 galaxy is very high in the infrared, with $L_{IR}$ estimated to
 be $2 \times 10^{11}$ times the solar luminosity.  This galaxy is
 among a set of 31 identified OH megamasers (Henkel and Wilson 1990).
 Henkel and Wilson state that as a class, this sample of objects
 ``probably consists of strongly interacting pairs of galaxies
 undergoing a starburst.'' Other than this, little information
 is available which would cause us to suspect that this galaxy
 is the source of the 320 EeV shower.

\par The direction of one prominent Seyfert galaxy, MCG 8-11-11, is
 close to the arrival direction of the 320 EeV shower (see Fig. 4.).
 It does not pass the cut $ z < 0.0125$, since it has $z=0.0205$.
 Its radio luminosity yields $R_L=0.7$, nearly passing the criterion
 for selecting potential sources.  Its X-ray and $\gamma$-ray
 luminosities greatly exceed its radio luminosity (Perotti et al. 1990)
 so that the observed total luminosity is not discouraging.  In the
 X-ray spectral range from 20-100 keV its luminosity is
 $4.6 \times 10^{44}$ ergs s$^{-1}$, while in the low energy gamma ray
 range from 0.09-3 MeV the measured luminosity is even higher
 ($7 \times 10^{46}$ ergs s$^{-1}$), according to Perotti et al (1981).

 \par Such a large luminosity would allow this source to be
considered as a potential source of the 320 EeV shower, in spite
of the very significant attenuation that is expected at that
distance.  There is a serious problem with the scenario involving
MCG 8-11-11, however, as described in the previous section.  The
problem is that a source spectrum at this distance which would
have a reasonable chance of producing a 320 EeV shower at Earth
would be expected to produce an essentially unattenuated flux of
10 EeV showers at Earth that would be too large to agree with
Fly's Eye data. Because of this problem, MCG 8-11-11 is not
considered to be a likely source of the 320 EeV shower unless the
assumptions in the argument of the previous section are violated.
For example, the lower energy ($\sim $10 EeV particles) might be
efficiently trapped near the source somehow or the detected
particle might somehow be immune to interactions with the cosmic
background radiation.

\par The search described so far has targeted candidate production
 sites near the arrival direction of the detected air shower.  If
the extragalactic magnetic field is considerably larger than
$10^{-9}$ G, however, and does not reverse its direction
frequently along particle trajectories, the deflection angle of
the primary particle can be large. To select potential sources of
primary protons or nuclei in this case, we only require the
sources to be powerful and nearby, with no cut on the source
direction.  With the redshift cut ($z<0.0125$), there are 24
candidates remaining in the Burbidge-Crowne catalog referred to
above.  It is of interest to discriminate between the candidates
on the basis of $R_L$.  In Table 1, candidates are listed which
have $R_L$ exceeding unity or radio fluxes at 408 MHz exceeding
10 Jy.  The values of $R_L$ have been used to rank the candidates
by radio luminosity at 408 MHz. The ranking gives M87, For A, and
Cen A as the top 3 choices. These may be outstanding choices as
source candidates in the ``high'' magnetic field model, but there
is no clearly favored candidate since the arrival direction is
not useful in discriminating between candidates in this case.

\begin {center} \begin{tabular}{|c|c|c|c|}
\hline
 Object & Redshift & 408 MHz flux& $ R_L$ \\
{} & Parameter, z & Jy& {} \\ \hline
M87&0.0043&510& 17. \\
For A&0.0063&177& 12. \\
Cen A&0.0016&2400& 11. \\
NGC 4261&0.0073&37& 3.5 \\
NGC 5090 &0.0110&10& 2.1 \\
NGC 4696 &0.0093&7& 1.1 \\
M84&0.0031&12& 0.21 \\
M82&0.0009&12& 0.017 \\ \hline
\end{tabular}
\end {center}
\begin{center}
\bf {Table 1: Nearby high flux radio galaxies}
\end{center}
\vspace{5. mm}

\par In the above cases, the primary particle was assumed to be
 deflected by the extragalactic magnetic field and limited in range.
 A more speculative proposal is that the primary particle was a
 currently unknown kind of neutral particle (or an already known
 neutral particle with unexpected high energy properties) which did
 not suffer energy losses in propagating intergalactic distances.
 In this case the source could be far away, but it should be powerful
 and it should be contained within the inner error box shown in
 Fig. 4. We can search for such a source in a sample of high radio
 luminosity sources.  A ``complete sample'' of powerful radio sources
 with $\delta \ge 10^\circ$ has been collected by Herbig and Readhead
 (1992).  There are 173 of those sources, and the expected number in
the solid angle of our 2$\sigma$ error box is 0.24.

	One of these sources falls within the inner error box in
Fig. 4.  It is 3C147.  The position of 3C147 is plotted in the
figure, where it can be seen that its position is actually within
a $1\sigma$ error box.  Its position might also be consistent
with the remarkably high energy shower detected at Yakutsk
(Efimov et al. 1991; Sommers 1993).  The total radio luminosity
of 3C147 is $7.9
\times 10^{44}$ $erg$ $s^{-1}$ (Herbig and Readhead, 1992).
It's x-ray luminosity is approximately the same (Zamorani et al.
1981).  There are 60 sources in the Herbig and Readhead catalog
with radio luminosities as high as this value.  Comparing the
radio flux at 2.5 GHz, 28 of the 173 sources in that catalog
are as bright as 3C147.

\par  Previously we noted the requirements pointed out by
 Hillas (1984) for extremely high energy accelerators to have large
 magnetic fields extending over large distance scales.  Such large
 scale fields, combined with large plasma densities, would produce
 very large Faraday rotations.  In a search for extremely high Faraday
 rotations, Kato et al. (1987) measured the Faraday rotation measures
 of 100 extragalactic sources which were considered to be candidates
 for having very large rotation measures. They found a value of
 $-1,510 \pm 50$ $rad$ $m^{-2}$ for 3C147, putting it in the top four
 sources in their candidate sample. (The other 96 sources did not give
 clear evidence for rotation measures with magnitudes exceeding
 500 $rad$ $m^{-2}$). This result may add some plausibility to the
 hypothesis that 3C147 could be the source of the primary
particle.

\par The searches described here have assumed that the 320 EeV
 particle was accelerated at or near a galaxy. If the source of the
 primary particle were a cosmic string, or some other non-galactic
 source, there is of course no catalog that can be searched. A cosmic
 string is intrinsically powerful, it could be nearby, and it could
 be in any direction.

\section{Magnetic bending and nearby radiogalaxies}

	The source search described in the previous section finds
no likely astrophysical source for a 320-EeV particle within the
distance limits described in section 2 and near the arrival
direction of that shower.  One possibility is that the detected
particle was a charged particle whose trajectory was bent through
a large angle en route from the source.  If the bending is due to
extragalactic magnetic fields, this means that the line-of-sight
distance to the source must be even shorter than the distance
limits of section 2.  It also means that the charged particle
would have been produced long before those photons which are
detected in contemporary studies of astrophysical sources.  An
additional 5 Mpc pathlength, for example, would mean that the
acceleration occurred 15 million years prior to the state of the
system as it is observed with photons.  Although there are no
strong radiogalaxy hot spots seen within 50 Mpc at the present
time, the situation may have been very different 15 million years
ago.  The synchrotron lifetime of radio hot spot electrons can be
less than the light travel time from the nucleus to a hot spot
(Bridle \& Perley 1984), less than a million years.  Moreover,
analyses of radiogalaxy populations indicate that stronger
radiogalaxies have shorter lifetimes, with the strongest sources having
lifetimes less than 10 million years (Schmidt 1966; van der Laan
\& Perola 1969).  Multiple double-lobe structures in
radiogalaxies have been interpreted as evidence for intermittent
activity on such time scales (Leahy \& Parma 1992).  Candidate sources for
the 320 EeV particle should therefore include such radiogalaxies
even if the lobes do not presently have strong hot spots.

	Cen A is a radiogalaxy which is only 3 Mpc (Hesser et al.
1984) or perhaps 5 Mpc (Burbidge \& Burbidge 1959) away.  It has
two sets of double radio lobes.  The size of the entire system is
as large as that of strong radiogalaxies like Cygnus A.  The
widely separated radio lobes can be attributed to much greater
activity in the past (Hey 1983).  The line of sight to Cen A makes an
angle of 136$^{\circ}$ with the arrival direction of the 320-EeV
shower, so Cen A is a candidate source only if large magnetic
deflection is plausible.

	Virgo A (M87) is in the Virgo cluster, at a distance of
13-26 Mpc.  It also has multiple lobe structure on large scales
(Kotanyi 1980), which could be indicative of greater activity in
the past.  The line of sight to Virgo A makes an angle of
87$^{\circ}$ with the arrival direction of the 320-EeV air
shower, so large magnetic deflection would be needed for it to be
the source also.

	Less powerful nearby radiogalaxies are M82 and M84.
M82 is a starburst galaxy which is only about 3.5 Mpc away.
It makes an
angle of 37$^{\circ}$ with the 320-EeV particle's arrival
direction.  The possibility that the primary cosmic ray is
a heavy nucleus accelerated at M82 is discussed later in this
section.

	To estimate the strength of magnetic fields required to
make any of the nearby radiogalaxies a viable candidate, one can
ask what uniform transverse field B$_T$ (in $\mu$G) is needed to
bend the trajectory of a particle of energy E (in EeV) and charge
Z through $\phi$ degrees in the course of a pathlength D (in
Mpc).  The Lorentz force produces angular change according to
\begin {equation} \phi = 5.3\times 10^{4}\ Z B_T D/E.
 \end{equation} For example, suppose Z=1
and E=320.  Then for $\phi$ to be a $90^{\circ}$ bend over a path
of D=10 Mpc requires B$_T=0.05\ \mu G$.  Scale sizes for regions
of uniform field direction might not be greater than 1 Mpc,
however.  For a proton trajectory to bend through 90$^{\circ}$ in
1 Mpc requires $B>0.5 \mu G.$ Similarly, $5 \mu G$ would be
needed over a region of 100 kpc dimension.  Ptushkin (1991) estimated
that extragalactic fields of 0.1 $\mu G$ are likely, with
directional coherence over distances of about 1 Mpc.  That is
slightly lower than what may be needed if the 320 EeV particle
was a proton whose trajectory was bent from a nearby radiogalaxy.

	The required extragalactic magnetic field strength can be
reduced by 1/Z if the detected particle were a nucleus with
charge Z.  Because of photonuclear disintegration, that requires
the source to be closer than if it were a proton.  For
example, an iron nucleus might lose half of its nucleons during a
transit of 6 Mpc and arrive with Z=13.  The required magnetic
field strengths are then reduced by more than an order of magnitude.

	Extragalactic fields are usually inferred observationally
from Faraday rotation measure studies, which require a
sufficiently high product of magnetic field (parallel to the line
of sight) times electron density integrated over the thickness of
the observed field region.  Even relatively strong magnetic
fields can go undetected because the electron density is not high
and/or the integration distance is not long.  Rotation measure
studies have succeeded in detecting magnetic fields in regions of
adequately high electron densities.  For example, a field
strength of $1.5 \mu G$ has been inferred for a 10 Mpc region
centered on the Virgo Cluster (Vallee 1990a).  A $2
\mu G$ field has been reported in the Coma cluster (Kim et al.
1990), and a field of 0.3-0.6 $\mu G$ apparently exists between
the Coma cluster and the Abell 1367 cluster over a distance of
approximately 40 Mpc (Kim et al. 1989).  The existence of fields of such
magnitude elsewhere suggests that the magnetic fields needed to
bend this particle's trajectory might be present.

	It is conceivable that cosmic rays produced in the Virgo
cluster drive a cluster wind, similar to the solar wind, which
might extend to the Local Group and beyond.  The magnetic field
strength falls like 1/r in such a wind.  The integrated magnetic
field energy in a wind which extends to a radius of 20 Mpc and
has a field strength of 0.1 $\mu G$ at 20 Mpc is not greater than
the field energy in the Virgo cluster's 1.5 $\mu G$ field if it
is uniform out to a radius of 5 Mpc (Vallee 1990a).  Active galactic nuclei
throughout the history of the Virgo cluster may have provided the
energy for building up the observed field and for powering the
hypothetical wind.

	Disordered magnetic fields may exist as a result of past
activity by galactic nuclei which are now dormant, even if there
is no cluster wind.  A portion of the prodigious energy released
by an AGN is converted to magnetic fields at giant radio lobes.
Fields of $\sim 100\ \mu G$ seem to envelope the lobes of Cyg A
(Dreher et al. 1987), for example.  Due to the high plasma
conductivity, the dissipation time for such magnetic fields is
long compared to the Hubble time (Parker 1979).  Relic radio lobes,
left from the early universe when quasars were common, may have a
high population density within the Virgo Supercluster, including
our own neighborhood.  They may act as magnetic scattering
centers.

	Stringent upper limits exist for a uniform magnetic field
of cosmological scale, but those limits do not preclude the
existence of smaller scale fields capable of bending
the trajectory of a 320 EeV particle.  Those studies are
based on Faraday rotation measures of distant quasars.  Although
a precise upper limit would require precise knowledge of the
universal electron density, a universal magnetic field greater
than $10^{-11}\ G$ is inconsistent with reasonable estimates for
the electron density (Vallee 1990b).  This result does not bear directly on
the interpretation of the 320 EeV particle's origin, because it
pertains only to a hypothetical field whose direction would be
constant over the observable universe.

	If this picture of the Fly's Eye particle originating at
a nearby radiogalaxy is correct, it is possible that the cosmic
ray spectrum above the ankle transition (Bird et al. 1993a) may
be dominated by particles from the same source.  If 320 EeV
protons are magnetically deflected through large angles, then
lower energy particles may be magnetically confined in a region
containing the source as well as our Galaxy.  The lower energy
particles may diffuse out more slowly, steepening the spectrum
from a source spectral index of 2 to a larger spectral index at
detection, as is presumed to happen with Galactic sources and
confinement below the spectrum's knee.  It is interesting that
the spectral index above the ankle transition has been measured
to be 2.7 (Bird et al. 1993a).

	The trajectory bending has so far been assumed to be
accomplished by extragalactic magnetic fields.  The possibility
of large-angle bending by the Galaxy's fields should also be
explored.  For B$_T=3\ \mu G$, Z=1, and E=320, equation 7 gives a
deflection of 0.5$^{\circ}$ per kpc.  The pathlength through the
magnetic disk should not be greater than about 4 kpc, so the
Galaxy's disk should not deflect such a proton more than about
2$^{\circ}$.  However, there is also the possibility of a
galactic wind field (Johnson \& Axford 1971; Jokipii \& Morfill
1985) extending perhaps to a radius of 0.3 Mpc.  Since the
B-field of the wind falls as 1/r with radius, the differential
form of equation 7< is needed: \begin{equation} d\phi = 5.3\times
10^4\ (Z/E) B_T(r) dr.\end{equation} Using $B_T=3 \mu G\times
(.01Mpc/r)$ and integrating from r=.01 Mpc to r=.3 Mpc yields an
angular deflection of $\phi = 17^{\circ}$.  That is still not
enough to make the 320 EeV particle's direction point toward an
interesting nearby radiogalaxy if it was a proton.  Large angle
bending in the Galaxy's wind is plausible, however, for atomic
nuclei.  In that case, the particle's path need not be much
longer than the straight line photon paths, so changes in the
source since its production should not be important.

	One interesting possibility is that the Fly's Eye
detected a nucleus which originated in M82 as an iron nucleus.
Although M82 is only 3.5 Mpc away, this scenario seems unlikely
unless the true energy of the particle was less than 320 EeV.
Nuclei disintegrate too rapidly at energies above 300 EeV.  The
picture can work, however, if we assume that the energy of
the detected particle was actually 230 EeV, which is within the
$1\sigma$ error bar on its energy.  As shown in Figure 5, a
differential spectrum of starting iron nuclei develops a pile-up
bump near 230 EeV after propagating 3.5 Mpc.  The mean mass in the
bump is $<$A$>$=38 and the mean charge is $<$Z$>$=18.  We can
therefore picture M82 as the source of an original iron nucleus
with an energy of $\sim$340 EeV which was detected at Earth as a
Z$ \sim$ 18 nucleus of energy $\sim$230 EeV.

	Even the magnetic disk of the Galaxy might then account
for a 37-degree deflection of the 230 EeV nucleus with Z=18.  For
$\phi=37^{\circ}$, $B_T=3\ \mu G$, and Z=18, equation 7 requires
only a pathlength of $D=0.0030\ Mpc = 3.0\ kpc$ through the
Galaxy's magnetic disk.  This is not unreasonable for this
particle, in view of its equatorial arrival direction
($b=10^{\circ}$).  The regular magnetic field of the
Galaxy is known to be parallel to the galactic plane and point
approximately toward $b=90^{\circ}$ (Rand \& Kulkarni 1989).  The
position of M82 ($b=41^{\circ},\ l=141^{\circ}$) could then be
consistent with this particle's arrival direction
($b=9.6^{\circ},\ l=163^{\circ}$) if it had that positive charge.
The atmospheric depth of maximum of the air shower is also
compatible with it being a mid-size nucleus.

Although M82 is not a strong radiogalaxy, it has been described
as the archetypal starburst galaxy (Fitt and Alexander 1993) and
as a prototype of superwind galaxies (Heckman et al. 1990).  More
than 40 discrete radio sources have been observed within the
central 700 pc of M82 (T.W.B. Muxlow et al. 1994) most of which
are supernova remnants which are probably only a few hundred
years old. If older remnants are considered, the number must be
very large.  Atomic nuclei might be accelerated to superhigh
energies through collisions with numerous expanding supernova
remnants (Axford 1991).  A bipolar magnetic field of perhaps 50
$\mu G$ is oriented perpendicular to the galactic plane (H.-P.
Reuter et al.  1992).  In a magnetic field of that strength, the
gyroradius of a 370 EeV iron nucleus would be only $\sim$300 pc.
There is also a kpc-scale bipolar wind which is emitted
perpendicular to the galactic plane, originating in the starburst
region (Bland and Tully 1988).  As in the model of Jokipii and
Morfill (1985), the termination shock of the galactic superwind
provides a possible site for the acceleration of superhigh energy
nuclei.  Shock acceleration could be especially efficient where
the superwind meets galactic winds from its neighbors, M81 and
NGC 3077.

\section{Discussion}

	There is no simple explanation for the 320 EeV cosmic ray
detected by the Fly's Eye.  The distance limits of section 2
imply that it originated within a radius of 50 Mpc.  Very powerful
radiogalaxies are the best candidate sources for such energetic
particles, but these radiogalaxies are more
than 100 Mpc away.  We have suggested that some nearby
radiogalaxies, such as Cen A or Virgo A, may have been much
stronger in the past when this particle was produced.  Because
it did not follow a straight line trajectory (there are no close
enough candidate sources near its arrival direction), it would
have been produced long before the photons with which we study
the nearby radiogalaxies.  The viability of this hypothesis
hinges on the existence of extragalactic magnetic fields which
are capable of bending its trajectory.  Because of the particle's
high magnetic rigidity, relatively strong fields are needed.  We
have argued that such strong fields are not excluded
observationally and may not be implausible.

	The problem of magnetic bending is less severe if the detected
particle was a highly charged nucleus.  It is then possible that
it was deflected significantly by the Galaxy's magnetic field.
As a possible scenario, we have noted that the starburst galaxy
M82 is only 37$^{\circ}$ from the particle's arrival direction,
and positioned so that a positively charged particle from its
direction would have been deflected toward the arrival direction
of this particle.  M82 is only about 3.5 Mpc away, so
photodisintegration would not be expected to erode too much of
its charge during transit, provided the energy at detection is
somewhat less than the best experimental estimate.  If this scenario is
correct, detections of other superhigh energy nuclei can be expected to
lie on a curve of directions determined by the position of M82
and the Galaxy's regular field and parametrized by magnetic rigidity.

	An intriguing Seyfert galaxy MCG 8-11-11 lies very close
to the arrival direction of the 320 EeV shower.  It is a powerful
source of low energy gamma rays.  It apparently has ample
luminosity to account for a detectable flux of superhigh energy
particles (even allowing for severe attenuation by photopion
production losses).  A strong flux has not been detected,
however, at somewhat lower energies where attenuation is not
important.

	A speculative idea is that the Fly's Eye detected a 320
EeV neutral particle which is immune to interactions with the
cosmic background radiation.  The quasar 3C147 becomes a
candidate source in that case. Its position is consistent with
the arrival direction of the air shower.  3C147 is a remarkable
radio source with intense hot spots whose radio luminosity is
more than 2000 times greater than either Cen A or Virgo A.
Luminosity is still an issue for 3C147, however, due to its great
distance.  Its redshift is z=0.545, implying a luminosity
distance (Weinberg 1972) of 2200 Mpc for Hubble constant $H_0=75$
 km s$^{-1}$ Mpc$^{-1}$.  Dividing the 320 EeV energy by the Fly's Eye's
exposure to 3C147 gives a time-averaged energy flux of 11
eV/cm$^2\cdot$s.  The implied luminosity at the source is then
$10^{46}$ erg/s.  This is a high luminosity, even for an active
galaxy, and it certainly exceeds the radio and x-ray luminosities
of 3C147.  It is a challenge to explain such a high luminosity in
secondary neutral particles of superhigh energy.

	A radically different possibility is that the 320 EeV
particle did not originate at any persistent astrophysical object
which would be found in a catalog search.  It need not have been
accelerated at all.  It could have been produced by the
annihilation of a topological defect (Aharonian et al. 1992),
e.g. a cosmic string (MacGibbon \& Brandenberger 1993).  The
absence of a close enough astrophysical source near the arrival
direction of this high rigidity particle can be construed as
evidence in favor of topological defect annihilations as the
sources of superhigh energy particles.  This explanation for the
320 EeV particle has been discussed by others (Sigl, Schramm, \&
Bhattacharjee 1994).

	The decay of a topological defect is presumed to result
in the production of X-particles with masses near the GUT energy
scale of $10^{24}$ eV.  Superhigh energy particles result from
their decays.  Gamma rays may be the dominant detectable
particles at superhigh energies, since they are produced
copiously by $\pi^0$ decays and also inverse Compton scattering
by electrons and positrons which result from charged pion decays.
The Fly's Eye air shower could have been initiated by a gamma
ray.  Above 100 EeV, the LPM effect (Landau \& Pomeranchuk 1953;
Migdal 1957; Mizumoto 1993) is expected to cause gamma ray air
showers to develop anomalously deep in the atmosphere.  The Fly's
Eye shower was not anomalously deep.  Its depth of maximum was
typical for a hadronic shower.  It might also be consistent,
however, with a gamma ray which converted to an electron pair and
initiated its electromagnetic cascade in the earth's
magnetosphere before reaching the atmosphere (McBreen \& Lambert
1981).  Such a shower would behave as a superposition of
lower-energy electromagnetic air showers.  Cascading in the
magnetosphere would be especially likely if this particular
shower were a gamma ray, because it arrived transverse (at
85$^{\circ}$) to the local geomagnetic field.

	In summary, the 320 EeV Fly's Eye shower would seem to
offer an excellent opportunity to identify a source of the
highest energy cosmic rays.  As shown in section 2, that cosmic
ray almost surely originated within a distance of 50 Mpc.
Because of its high magnetic rigidity, one might expect that its
arrival direction should point approximately toward its source.
Unfortunately, the arrival direction of this shower is not near
the direction of any close enough astrophysical object known to
us as a likely acceleration site for such energetic particles.
Identifying the source of this particle constitutes an important
challenge in astroparticle physics.

\end{large}

\section{Acknowledgements}
\begin{large}
   We thank E.C. Loh and H.Y. Dai for many ideas.  We are also
grateful to D.N. Schramm for discussions in Utah and to W.A.
Christiansen for helpful telephone conversations.  This research
has made use of the NASA/IPAC Extragalactic Database (NED) which
is operated by the Jet Propulsion Laboratory, California
Institute of Technology, under contract with the National
Aeronautics and Space Administration. This work has been
supported by the National Science Foundation (grant
PHY-91-00221).
\end{large}
{}~~\\
{}~~\\

\newpage
\noindent{\Large \bf References}
{}~~\\
\begin{large}

\noindent\hangindent=20pt Aharonian, F. A., Bhattacharjee, P., \&
Schramm, D. N. 1992, Phys. Rev. D, 46, 4188

\noindent\hangindent=20pt Axford, W. I. 1991, in Astrophysical
Aspects of the Most Energetic Cosmic Rays, ed. M. Nagano \&
F. Takahara (Singapore: World Scientific), 406

\noindent\hangindent=20pt Baltrusaitis, R. M. et al. 1985, Phys.
Rev. D, 31, 2192

\noindent\hangindent=20pt Bird, D. J. et al. 1993a, Phys. Rev. Lett., 71,
3401

\noindent\hangindent=20pt Bird, D. J. et al. 1993b, {\em Proc. 23rd Int.
Cosmic Ray Conf. (Calgary)} 2, 51

\noindent\hangindent=20pt Bird, D. J. et al. 1994, ApJ, submitted

\noindent Bland, J.\&Tully, R. B. 1988, Nature, 334, 43

\noindent\hangindent=20pt Blumenthal, G. R. 1970, Phys. Rev. D, 1,
1596

\noindent\hangindent=20pt Bridle, A. H. \& Perley, R. A. 1984,
Ann. Rev. Astron. Astrophys., 22, 319

\noindent\hangindent=20pt Burbidge, E. M. \& Burbidge, G. 1959,
ApJ, 129, 271

\noindent Burbidge, G.\& Crowne, A. H. 1979, ApJ, 40, 583

\noindent\hangindent=20pt Chi, X., Szabelski, J., Vahia, M. N.,
Wdowczyk, J., \& Wolfendale, A. W. 1992, J. Phys. G: Nucl. Part.
Phys., 18, 539

\noindent\hangindent=20pt Clark,T.A., Brown, L. W. \&
Alexander, J. K. 1970, Nature, 228, 847

\noindent\hangindent=20pt Clarke, D. A., Burns, J. O., \& Norman,
M. L. 1992, ApJ, 395, 444

\noindent\hangindent=20pt De Jager, O. C., Stecker, F. W., \& Salamon, M. H.
1994
ApJ, in press

\noindent\hangindent=20pt de Vaucouleurs G. 1993, ApJ, 415, 10

\noindent\hangindent=20pt Domokos, G. \& Kovesi-Domokos, S. 1988,
Phys. Rev. D, 38, 2833

\noindent\hangindent=20pt Domokos, G. \& Nussinov, S. 1987,
Phys. Lett. B, 187, 372

\noindent\hangindent=20pt Dreher, J. W., Carilli, C. L., \&
Perley, R. A. 1987, ApJ, 316, 611

\noindent\hangindent=20pt Efimov, N. N., Egorov, T.A., Glushkov, A.V.,
Pravdin, M. I. \& Sleptsov, I. Ye. 1991, Astrophysical
Aspects of the Most Energetic Cosmic Rays, ed. M. Nagano \&
F. Takahara (Singapore: World Scientific), 20

\noindent\hangindent=20pt Emerson, B. L. 1992, Ph.D. thesis,
Univ. of Utah

\noindent Fitt, A. J.\& Alexander, P. 1993, MNRAS, 261, 445

\noindent\hangindent=20pt Gould, R. J. \& Schreder, G. P. 1967,
Phys. Rev., 155, 1408

\noindent Greisen, K. 1966, Phys. Rev. Letters, 16, 748

\noindent\hangindent=20pt Halzen, F., Protheroe, R. J.,Stanev, T.,\&
 Vankov, H. P. 1990, Phys. Rev. D, 41, 342

\noindent Heckman, T. H.,Armus L. \& Miley, G. K. 1990, ApJS, 74, 833

\noindent\hangindent=20pt Helou, G., Madore, B. F., Schmitz, M. D.,
 Bicay, M. D., Wu, X. \& Bennet, J. 1991, in Databases and On-Line
 Data in Astronomy, ed. D. Egret \& M. Albrecht (Dordrecht: Kluwer), 89

\noindent Henkel, C. \& Wilson, T. L. 1990, A\&A, 229, 431

\noindent Herbig, T. \& Readhead, C. S. 1992, ApJS, 81, 83

\noindent\hangindent=20pt Hesser, J. E., Harris, H.C., van den Bergh,
 S. \& Harris, G.L.H. 1984, ApJ, 276, 491

\noindent\hangindent=20pt Hey, J. S. 1983, The Radio Universe
(3rd ed. New York: Pergamon)

\noindent\hangindent=20pt Hill, C. T. \& Schramm, D. N. 1985,
Phys. Rev. D, 31, 564

\noindent\hangindent=20pt Hillas, A. M. 1984, ARA\&R, 22, 425

\noindent\hangindent=20pt Johnson, H. E. \& Axford, W. I. 1971,
 ApJ, 165, 381

\noindent Jokipii, J .R.\& Morfill, G. E. 1985, ApJ, 290, L1

\noindent Jonsson, G. G.\& Lindgren, K. 1973, Physica Scripta, 7, 49

\noindent Jonsson, G. G.\& Lindgren, K. 1977, Physica Scripta, 15, 308

\noindent Kato, T., Tabara, H., Inoue, M. \& Aizu, K 1987, Nature, 329, 223

\noindent\hangindent=20pt Kim, K.-T. et al. 1989, Nature, 341,
 720

\noindent\hangindent=20pt Kim, K.-T. et al. 1990, ApJ, 355, 29

\noindent\hangindent=20pt Kotanyi, C. 1980, A\&A, 83, 245

\noindent\hangindent=20pt Landau, L. \& Pomeranchuk, I. 1953, Dokl. Akad.
Nauk (USSR), 92, 535 and 735

\noindent\hangindent=20pt Leahy, J. P. \& Parma, P. 1992 in
Extragalactic Radio Sources -- From Beams to Jets, ed. J. Roland,
H. Sol, \& G. Pelletier (Cambridge: Cambridge Univ. Press), 307

\noindent\hangindent=20pt Linsley, J. 1963, Phys. Rev. Lett., 10, 146

\noindent\hangindent=20pt MacGibbon, J. H. \& Brandenberger, R.
H. 1993, Phys. Rev. D, 47, 2283

\noindent\hangindent=20pt Madore, B. F., Freedman, W. L., \& Lee, M. G.
 1993, AJ, 106, 6

\noindent\hangindent=20pt McBreen, B. \& Lambert, C. J. 1981,
Proc. 17th International Cosmic Ray Conf. (Paris), 6, 70

\noindent\hangindent=20pt Migdal, A. B. 1957,JETP (USSR), 32, 633

\noindent\hangindent=20pt Mizumoto, Y. 1993, Proc. Tokyo Workshop on
Techniques for the Study of Extremely High Energy Cosmic Rays,
(Tokyo, Japan), ed. M. Nagano, 194

\noindent\hangindent=20pt Muxlow, T. W. B., Pedlar, A.,
 Wilkinson, P. N., Axon, D. J., Sanders, E. M. \&
 de Bruyn, A. G. 1994, MNRAS, 266, 455

\noindent\hangindent=20pt Parker, E. N. 1979, Cosmical Magnetic
Fields (Oxford:Clarendon Press)

\noindent\hangindent=20pt Perotti, F., Della Ventura, A., Villa, G., Di Cocco,
G.,
Butler, R. C., Carter, J. N.,\& Dean, A. J. 1981, Nature, 290, 133

\noindent\hangindent=20pt Perotti, F., Bassani, L., Bazzano, A.,
 Court, A. J., Dean, A. J., Lewis, R. A., Maggioli, P., Quadrini, M.,
 Stephen, J. B. \& Ubertini, P. 1990, A\&A, 234, 106

\noindent Puget, J. L., Stecker, F. W.,\& Bredekamp, J. H. 1976,
 ApJ, 205, 638

\noindent\hangindent=20pt Ptushkin, V. S. 1991, Astrophysical
Aspects of the Most Energetic Cosmic Rays, ed. M. Nagano \&
F. Takahara (Singapore:World Scientific), 112

\noindent\hangindent=20pt Rachen, J. P. \& Biermann, P. L. 1993,
 A\&A, 272, 161

\noindent\hangindent=30pt Rand, R. J. \& Kulkarni, S. R. 1989,
 ApJ, 343, 760

\noindent Reuter, H.-P., Klein, U., Lesch, H., Wielebinski, R.\&
 Kronberg, P. P. 1992, A\&A, 256, 10

\noindent Rudstam, G. 1966, Zs. f. Natursforschung, 21a, 1027

\noindent\hangindent=20pt Schmidt, M. 1966, ApJ, 146, 7

\noindent\hangindent=20pt Sigl, G., Schramm, D. N. \&
Bhattacharjee, P. 1994, Astroparticle Physics, submitted

\noindent\hangindent=20pt Sommers, P. 1993, Proc. Tokyo Workshop on
Techniques for the Study of Extremely High Energy Cosmic Rays,
(Tokyo, Japan), ed. M. Nagano, 23

\noindent Silberberg, R.\& Tsao, C. H. 1973a, ApJS, 25, 315

\noindent Silberberg, R.\& Tsao, C. H. 1973b, ApJS, 25, 335

\noindent Stecker, F. W. 1969, Phys. Rev., 180, 1264

\noindent\hangindent=20pt Stecker, F. W., Done, C., Salamon, M. H.
 \& Sommers, P. 1991, Phys. Rev. Lett., 66, 2697

\noindent\hangindent=20pt Vallee, J. P. 1990a, AJ, 99, 459

\noindent\hangindent=20pt Vallee, J. P. 1990b, ApJ, 360, 1

\noindent\hangindent=20pt van den Bergh, S. 1992, PASP, 104, 861

\noindent\hangindent=20pt van der Laan, H. \& Perola, G. C. 1969,
A\&A 3, 468

\noindent\hangindent=20pt Weinberg, S. 1972, Gravitation and
Cosmology (New York:Wiley)

\noindent\hangindent=20pt Wdowczyk, J., Tkaczyk, W. \&
Wolfendale, A. W. 1972, J. Phys. A, 5, 1419

\noindent\hangindent=20pt Yoshida, S. 1994, Astroparticle
 Physics, submitted

\noindent\hangindent=20pt Zamorani, G. et al. 1981, ApJ, 245, 357

\noindent Zatsepin, G. T.\& Kuz'min, V. A. 1966, JETP Lett., 4, 78

\end{large}
\newpage

{}~~\\
{}~~\\
{\Large \bf Figure Captions}\\
\begin{large}
{}~~\\
 {\bf Figure 1.}  Integral flux reduction factors as a function of
 energy, after a spectrum with a differential spectral index of 2.5
 has been propagated the indicated distances. Results are given for
 the following initial particles:
 (a) protons, (b) carbon nuclei, and (c) iron nuclei. \\
{}~~\\
 {\bf Figure 2.}  The dependence of the integral flux reduction
 factors on path length for cosmic rays with energies greater than
 320 EeV. Spectra with differential spectral indices of 2., 2.5,
 and 3. are indicated by solid, dashed, and dotted curves,
 respectively. Results are given for (a) protons, (b) carbon nuclei,
 and (c) iron nuclei. \\
{}~~\\
 {\bf Figure 3.} Probability of sources at various distances,
 considering only propagation effects (see discussion in text).
 Spectra with differential spectral indices of 2.,2.5, and 3. are
 indicated by solid, dashed, and dotted curves, respectively.
 Results are given for (a) protons, (b) carbon nuclei, and
 (c) iron nuclei. \\
{}~~\\
  {\bf Figure 4.}  The source search region is shown by the outer
 boundary. It is defined by angular distances of $10^\circ$ from the
 $2 \sigma$ error box shown by the dotted lines. The best fit shower
 direction is shown by the cross. The circle and square represent
 the directions of 3C 147 and MCG 8-11-11, respectively. The triangle
 shows the direction of NGC 1569 (also known as Arp 210.) The + symbol
 is in the direction of UGC 03351, also known as MCG 10-09-02.
 See text for details. \\
{}~~\\
 {\bf Figure 5.} Ratio of surviving differential flux to source flux
 for primary iron nuclei as a function of the detected energy after a
 pathlength of 3.5 Mpc.  The differential spectral index was chosen to
 be $\gamma =$2.5. A ``pile-up'' is observed near the shower energy
 (230 EeV, marked by a dotted line) proposed in a scenario involving
 acceleration of the cosmic ray near M82. (See text.)
\end{large}
\end{document}